\documentclass[preprint,12pt]{elsarticle}

\usepackage{amssymb}
\usepackage{color}
\usepackage[english]{babel}

\journal{Physica A}

\begin{document}

\begin{frontmatter}



\title{When {\it I cut, you choose\/} method implies intransitivity}


\author{Marcin Makowski\corref{cor1}}
\ead{makowski.m@gmail.com}
\author{Edward W. Piotrowski\corref{cor2}}
\ead{qmgames@gmail.com}
\address{Institute of Mathematics,
                University of Bia\l{}ystok\\
								ul. Akademicka 2, 15-267 Bia\l{}ystok, POLAND\\ phone: +48 85 745 7544}

\begin{abstract}
There is a common belief that humans and many animals follow transitive inference (choosing A over C on the basis of knowing that A is better than B and B is better than C). Transitivity seems to be the essence of rational choice. We present a theoretical model of a repeated game in which the players make a choice between three goods (e.g. food). The rules of the game refer to the simple procedure of fair division among two players, known as the ``I cut, you choose'' mechanism which has been widely discussed in the literature. In this game one of the players has to make intransitive choices in order to achieve the optimal result (for him/her and his/her co-player). The point is that an intransitive choice can be rational. Previously, an increase in the significance of intransitive strategies was achieved by referring to models of quantum games. We show that \textit{relevant intransitive strategies} also appear in the classic description of decision algorithms.
\end{abstract}

\begin{keyword}  intransitivity \sep relevant intransitive strategies \sep repeated game



\end{keyword}

\end{frontmatter}


\section{Introduction}
In this work we analyse a simple model of a repeated game of choice. This is an interesting modification of our earlier work \cite{r1,r2,r2n,En,El}.  To illustrate the problem we refer to the history of two players - cats (older and younger) that are making a choice between three types of food, no.~0, no.~1, and no.~2 (e.g. milk, meat and fish) in the following way: at the beginning of each stage the older cat rejects one of the foods. Then the younger cat selects and consumes one food from the other two.  The food that was rejected by the younger cat is consumed by the older one. The same procedure is repeated in each iteration of the game. Let us assume that both players tend to obtain an equal contribution of each of the foods in their diet (i.e. one third for food no.~0, no.~1 and no.~2, respectively). The optimum consists in not distinguishing any of the three foods. Each of the foods is equally important to each of the players.

The game quoted above is interesting for at least two reasons. Firstly, because the course of the game refers to a procedure that has been widely discussed in the literature. It is the simple procedure of fair (and envy-free) division among two players, known as the ``I cut, you choose'' mechanism \cite{Brams}. According to this rule one of the players makes a certain division, e.g. splits the pie into two parts, whereas the second player chooses one of the parts for him or herself. The second part of the pie falls to the player who made the cut. Therefore, the player who has made the division will certainly get at least half of the goods according to his/her own valuation. On the other hand, the player who makes the choice takes the part he/she prefers, so he/she can decide to take over that part which seems to make up at least half of the pie, as measured by his/her own assessment. A similar procedure is present in our game, where the older cat acts as the cutting player and the younger cat as the choosing one. The very procedure may be treated as a specific mechanism designed to be resistant to bluffs and cheating.

The second reason why the offered game appears interesting is its relation with the notion of transitivity-intransitivity of preferences. We can observe how the players' execution of the procedure of fair division  ``I cut, you chose'' influences the type of strategy (transitive or intransitive) they use. It appears that intransitive orders (considered by many researchers to be paradoxical) are a natural consequence of a conjunction of assumptions: a desire to obtain a variety of foods and a fair division of the available set of foods. A rejection of intransitivity means that one of these two requirements will have to be revised.

\section{A brief introduction to intransitivity}
\begin{quote} \textsl{I remember that at the age of eight or nine I tried to rate the fruits I liked in order of ``goodness''. I tried to say that a pear was better than an apple, which was better than a plum, which was better than an orange, until I discovered to my consternation that the relation was not transitive—namely, plums could be better than nuts which were better than apples, but apples were better than plums. I had fallen into a vicious circle, and this perplexed me at that age. Mathematicians' ratings are something like this.} 
\begin{flushright}
\protect{Stanisław Ulam \cite{Ulam}}
\end{flushright}
\end{quote}

The relation $\succ$, defined on elements of a certain set, is called \emph{transitive} if $A\succ C$ follows from the fact that $A\succ B$ and $B\succ C$ for any three elements  $A$, $B$, $C$. If this condition is not satisfied then the relation is called \emph{intransitive} (not transitive). 

The principle of the transitivity of preference is one of the basic assumptions of choice theory. It is often identified with the rationality of the players. The adoption of this assumption is probably a consequence of transitive inference. This ability is developed during childhood, i.e. at the age of four or five \cite{Pa}. It allows children to reason that if  \textbf{A} is bigger than \textbf{B}, and \textbf{B} is bigger than \textbf{C}, then \textbf{A} is also bigger than \textbf{C}. According to Jean Piaget, understanding transitive inference leads to the formation of skills of measuring, organising elements and deductive thinking in childhood \cite{Pa}. Transitive inference saves a large amount of time and energy in daily decision-making. This reasoning has been confirmed in some animal species \cite{r33,c,cc} (though there is no evidence that animals use it consciously). One of the main arguments put forward by many experts that proves the irrationality of preferences which violate transitivity is the so-called ``money pump'' \cite{tull,r23}. According to this argument, anyone who exhibited intransitive preferences would be punished by losing his/her entire wealth. On the other hand, some generalisations of utility theory have been considered which dispense of the transitivity assumption \cite{tw}. 
 
Despite the fact that intransitivity appears to be contrary to our intuition, life provides many examples of intransitive orders. There is mounting evidence based on empirical observations of choice behaviour that people do not have preferences that are consistent with transitivity \cite{T}. This suggests that transitivity does not describe people's preferences well. Rivalry between species may be intransitive \cite{k}; for example, among  fungi, Phallus impudicus replaces Megacollybia platyphylla, M.~platyphylla replaces Psathyrella hydrophilum, but P.~hydrophilum replaces P.~impudicus \cite{r15}. Similarly, an experiment can be found which presents the behavior of bees making intransitive choices between flowers \cite{r16}. The best known and socially significant example of intransitivity is Condorcet's voting paradox \cite{p}, in which collective preferences can be intransitive even if the preferences of individual voters are not (this means that majority preferences can be in conflict with each other). Consideration of this paradox led Arrow (in the 20th century) to prove the theorem that there is no procedure of successful choice that would meet a democratic assumption \cite{r24}. Interesting examples of intransitivity are provided by probability models (Efron's dice \cite{r34}, Penney's game \cite{r35}). Intransitivity models appear in many sciences (seemingly distant) including psychology \cite{T}, philosophy \cite{phy}, operations research \cite{ope}, coevolution \cite{co}, thermodynamics \cite{Ki},  quantum theory \cite{r2}, logic \cite{H}. 
\section{Details of the model}
Let us assume that cats are always offered the same three types of food (no. 0, no. 1 and no. 2).
The first player, \textsl{older cat}, chooses and rejects one of the foods.  Then the second player, \textsl{younger cat}, selects and consumes one of the remaining two foods. The older cat eats the food that is left.  Let us also assume that the cats will never refrain from making the choice of food.

The behaviour of the first player is limited to one choice of food to be eliminated of the three, so that the strategy can be described by the point $(P_0,P_1,P_2)$ of a three dimensional simplex, where $P_i$ denotes the frequency of the choice of a respective food (numbered by $i$) to be thrown away. 

The younger cat chooses between the two foods that, are left. Number $P(C_{k} | B_{j})$ denotes the probability of choosing (by the younger cat) food number $k$ when the offered pair of dishes does not contain food number $j$. Six conditional probabilities $P(C_{k} | B_{j})$, $k,j=0,1,2$, $k\neq j$  determine the behaviour of the younger cat with respect to the proposed it in pairs food portions -- the frequency of their offer. 

The younger cat's strategy space is  a three-dimensional cube of three independent, conditional probabilities (we explicitly obtain the other three from the condition of normalisation of the probability measure to unity). A diagram of the game is shown in Fig. 1.
\begin{figure}[h]
\centering
\includegraphics[width=11.25cm,height=8.25cm]{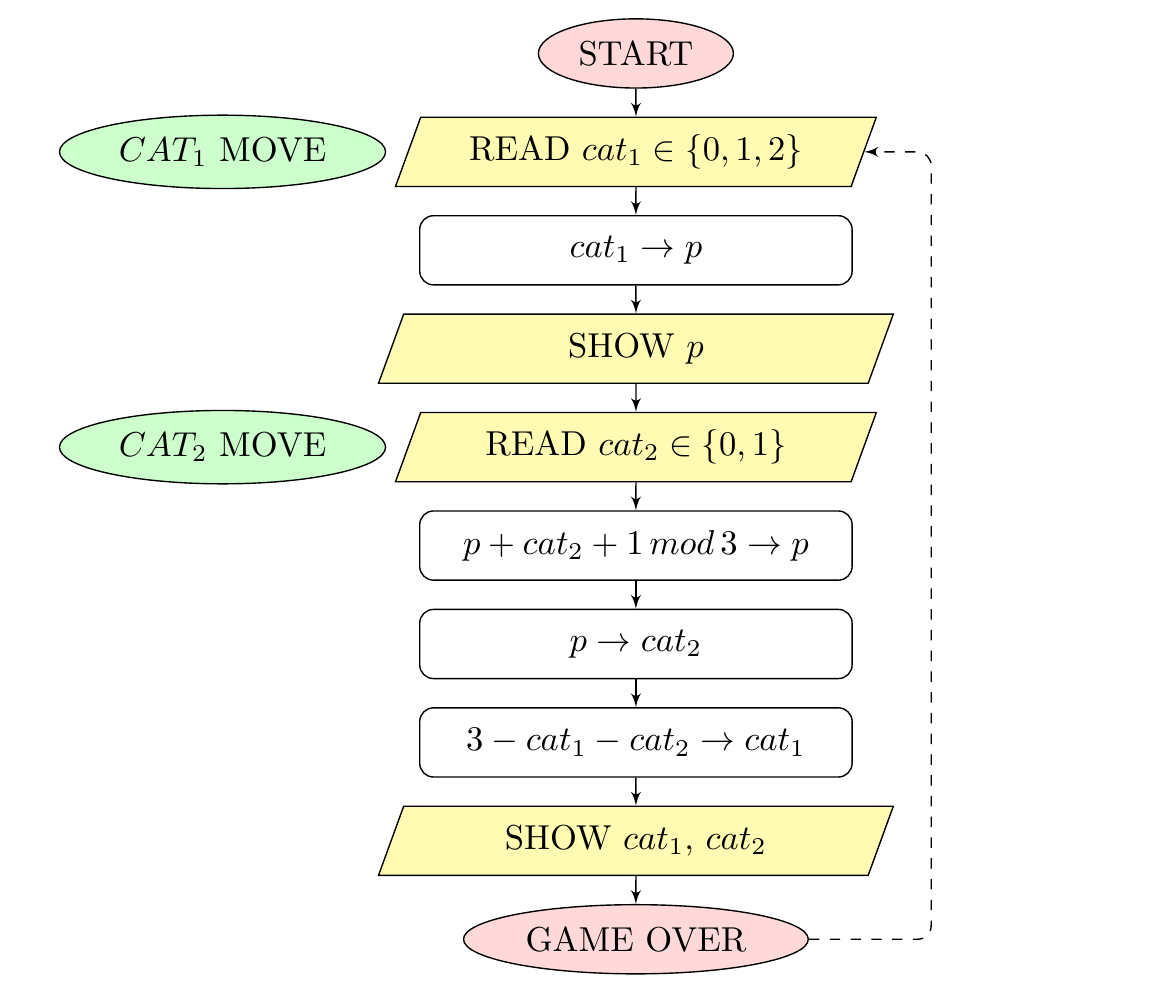}
\caption{Each iteration of the game. Symbol \textsl{mod n} is the remainder of the division by \textsl{n}. Older and the younger cat -- respectively $cat_1$ i $cat_2$.}\label{schem1}
\end{figure}

Let $\lambda_k$ and $\omega_k$ denot the frequencies of appearance of the particular foods in the older and the younger cats diet, respectively.
 For the older cat, they are following:
\begin{eqnarray}\label{mlodszy}
\lambda_0&=&P(C_{2} | B_{1})P_1+P(C_{1} | B_{2})P_2,\nonumber\\
\lambda_1&=&P(C_{0} | B_{2})P_2+P(C_{2} | B_{0})P_0,\\
\lambda_2&=&P(C_{1} | B_{0})P_0+P(C_{0} | B_{1})P_1\,,\nonumber
\end{eqnarray}
and for the younger cat we get:
\begin{eqnarray}\label{starszy}
\omega_0&=&P(C_{0} | B_{1})P_1+P(C_{0} | B_{2})P_2,\nonumber\\
\omega_1&=&P(C_{1} | B_{0})P_0+P(C_{1} | B_{2})P_2,\\
\omega_2&=&P(C_{2} | B_{0})P_0+P(C_{2} | B_{1})P_1.\nonumber
\end{eqnarray}
The portions of food are equally attractive to them and necessary, because they contain all components to maintain good health. So the most valuable way to select foods corresponds to:
\begin{eqnarray}\label{mlo}
\lambda_0=\lambda_1=\lambda_2&=&\frac{1}{3}\,,\\\label{str}
\omega_0=\omega_1=\omega_2&=&\frac{1}{3}\,. 
\end{eqnarray}
To simplify the notation of the equations we introduce three parameters $t_i\in [-1,1]$, $i=0,1,2$:
\begin{eqnarray}\label{param}
P(C_2|B_0)=\frac{1+t_0}{2},&&  P(C_1|B_0)=\frac{1-t_0}{2},\nonumber \\
P(C_0|B_1)=\frac{1+t_1}{2},&&  P(C_2|B_1)=\frac{1-t_1}{2}, \\
P(C_1|B_2)=\frac{1+t_2}{2},&&  P(C_0|B_2)=\frac{1-t_2}{2}\,.\nonumber  
	\end{eqnarray}
Then the equations (\ref{mlo}) and (\ref{str}) take the form:
\begin{eqnarray}\label{1}
-t_1 P_1+t_2 P_2=2/3-(P_1+P_2)\,,\nonumber\\
-t_2 P_2+t_0 P_0=2/3-(P_0+P_2)\,,\\
-t_0 P_0+t_1 P_1=2/3-(P_0+P_1)\,.\nonumber
\\\nonumber\\\label{2}
t_1 P_1-t_2 P_2=2/3-(P_1+P_2)\,,\nonumber\\
t_2 P_2-t_0 P_0=2/3-(P_0+P_2)\,,\\
t_0 P_0-t_1 P_1=2/3-(P_0+P_1)\,.\nonumber
\end{eqnarray}	
By adding the appropriate equations from (\ref{1}) and (\ref{2}) we obtain
\begin{displaymath}
P_k=\frac{1}{3}\,,\,\,\,\,\,\, \,\,\,\,\, k=0,1,2, 
\end{displaymath}
while by subtracting we get
\begin{displaymath}
\frac{t_m}{t_n}=\frac{P_n }{P_m}=1\,.
\end{displaymath}
 The only strategies that meet both conditions (\ref{mlo}) and (\ref{str})  is the family corresponding to $P_k = \frac {1}{3}$, for $ k = 0,1,2 $ (the older cat strategy) and six conditional probabilities (the younger cat strategy):
\begin{eqnarray}\label{opt}
P(C_2|B_0)= P(C_0|B_1)=P(C_1|B_2)=\frac{1+t}{2}, \nonumber \\
P(C_1|B_0)= P(C_2|B_1)=P(C_0|B_2)=\frac{1-t}{2}, 
	\end{eqnarray}
where $t\in [-1,1]$\,.
	Let's look at a set of the younger cat's strategies (their type: transitive or intransitive).
	
	We say that a player prefers food no. 1 to food no. 0 ($1\succ 0$) when he/she is willing to choose food no. 1 more often than food no. 0 from the offered pair $(0,1)$ ($P(C_1|B_2)>P(C_0|B_2)$). The situation corresponds to an intransitive choice if one of the following two conditions is satisfied:

	\vspace{0.1cm}
\begin{enumerate}
 \item $P(C_0|B_2)<P(C_1|B_2)$, $P(C_1|B_0)<P(C_2|B_0)$, $P(C_2|B_1)<P(C_0|B_1)$\,,
 \vspace{0.3cm}
 \item $P(C_0|B_2)>P(C_1|B_2)$, $P(C_1|B_0)>P(C_2|B_0)$, $P(C_2|B_1)>P(C_0|B_1)$\,. 
\end{enumerate}
Note that condition $2.$ is fulfilled if $t\in[-1,0)$, while for condition $1.$ to be satisfied it is necessary to put $t\in(0,1]$. This means that if the older cat plays $P_k=\frac{1}{3}$, for $k = 0,1,2$, then the younger cat must make intransitive choices (!) in order to meet conditions (\ref{1}) and (\ref{2}) (an extreme case for $ t = 0 $ can be classified as intransitive indifference strategy).

\section{Election interpretation}
The model discussed above may be fruitfully interpreted from an elective point of view by referring to the procedure of two-phase elections which is well known in many countries. Let us assume that there are three candidates running in the election. The first phase eliminates one of the candidates, and the final choice between the remaining two candidates is made during the second phase. The choosing player in both phases is the same, i.e. the society. We look at this player from the standpoint of the passing of time (between the first and second phase). However, during the first phase of the election the choice (the rejection of one candidate) is made by the older society, whereas, during the second phase the younger society.\footnote{The adjectives ``older'' and ``younger'' do not refer here to the different players; they denote the same player making a choice at various times (the first and second phase).} The assumptions of the game discussed in the paper adhere to a certain specific situation, when noone of the candidates enjoys advantage over the others. Any of them has equal chance to win in the second phase. Any of them has equal chance to loose in the second phase. This matches the condition which requires that the probability of rejection of any candidate in the second phase (in the previous model it was the frequency of foods falling to the older cat) was equal to 1/3. It appears that when the collective elector i.e. the society is indecisive (does not favour any of the candidates) then its preference during the second phase is always intransitive. This result is compatible with intuition. If the society is indecisive then the election is non-conclusive just as intransitive preferences are considered to be.  
\section{Conclusion}
In papers \cite{r1,r2,r2n,En,El} we considered (in the context of the intransitivity of strategy) a game which has become an inspiration for our discussion here. The game's participant was a cat that made its choice from among three foods. In these models the problem basically concerned only one player, whereas the second player (so-called Nature) was not interested in the result of the game. This article presents a situation where both players strive to achieve a specific result. Additionally, the players perform sequential steps according to the procedure derived from the simplest method of fair division ``I cut, you choose''. It appears that under such conditions the dividing player (the older cat) does not prefer any of the three foods and rejects any of them with a probability of 1/3. It is natural to assume that the share of each of the foods in the cat's diet should be equal(condition (\ref{mlo})). It is surprising that the choosing player (the younger cat) may only use intransitive strategies. Therefore, an individual who is convinced that it is only acceptable to make choices based on a transitive order would not be able to fulfil the assumptions of the game presented here. Within our model the intransitive strategies have managed to entirely rule out the transitive ones. These are the so-called \textit{relevant intransitive strategies} \cite{El}, i.e. strategies for which one cannot indicate transitive strategies with identical consequences. Previously, an increase in the significance of intransitive strategies was achieved by referring to models of quantum games \cite{r2}  (see also \cite{rQ, rT}).  Let us note that this fundamental qualitative change (compared to the model with one cat and Nature) will be achieved when the requirement to use the food that was not chosen by the younger cat is introduced. This implies associations concerning the management of waste and the closing of the ecological cycle. Intransitive loops are considered in research studies on ecosystem models \cite{r40,r41}.
In this article we analysed the classical model in which there is a complete domination of intransitive strategies. We found the relation of this result with the method  ``I cut, you choose'' to be particularly interesting because the problem of fair division is an allegory of many substantial decision-related issues in our everyday lives (negotiations of agreements, division of property, etc.). Thus it comes as no surprise that researchers from various fields of science present their interest in this issue. Intransitive orders are, perhaps, not as paradoxical as they seem to be at first sight. The example of the game analysed in the present paper demonstrates that it might be a very promising direction of research to look for relations between procedures of fair division and the notion of intransitivity. This may result in many noteworthy observations and conclusions in the future.

This work is inspired by the thoughts of Hugo Steinhaus, who described experiments with cats in his diary \cite{ST}. It turns out that a cat, when facing the choice between fish, meat and milk, prefers fish to meat, meat to milk, and milk to fish! Steinhaus argued that intransitive preference protects the cat from the fate of Buridan's ass. It is worth mentioning that Steinhaus is also one of the leading promoters of fair distribution problems \cite{ST1}.
\section*{Acknowledgments}
This work has been supported by the project \textbf{Quantum
games: theory and implementations financed by the National Science
Center} under the contract no \textbf{DEC-2011/01/B/ST6/07197}.
\label{}
\section*{References}




\end{document}